\documentclass[onecolumn,12pt,amssymb,amsmath,preprintnumbers,nofootinbib,superscriptaddress,aps]{revtex4-1}
\usepackage{lineno}
\usepackage{bm,color,xcolor}
\usepackage{slashed} 
\usepackage{graphicx}
\usepackage{multirow}
\usepackage{comment}
\usepackage{mathtools}
\usepackage{appendix}
\usepackage[colorlinks=true
,urlcolor=blue
,anchorcolor=blue
,citecolor=blue 
,filecolor=blue
,linkcolor=blue
,menucolor=blue
,linktocpage=true
,pdfproducer=medialab
,pdfa=true
]{hyperref}

\usepackage{xparse}
\usepackage[all]{hypcap}
\usepackage{cleveref,cancel,dsfont}
\usepackage{physics}
\usepackage[com pat = 1.1.0]{tikz-feynman}

\let\Tr\relax

\DeclareMathOperator{\Tr}{Tr}

\newcommand{\Lag}[1][]{\mathcal{L}_{\text{#1}}}
\newcommand{\defeq}{\equiv}

\newcommand{\Z}[1][]{\mathbb{Z}_{#1}}

\let\deltafunc\delta
\DeclareDocumentCommand\delta{}{\trigbraces{\deltafunc}}

\begin{document}

\title{\bf \Large Accidental Suppression of Wilson Coefficients in Higgs Coupling}

\author{Yunjia Bao}
\email{yunjia.bao@uchicago.edu}
\affiliation{Department of Physics, University of Chicago, Chicago, IL 60637, USA}
\affiliation{Enrico Fermi Institute, University of Chicago, Chicago, IL 60637, USA}
\affiliation{Kavli Institute for Cosmological Physics,
University of Chicago, Chicago, IL 60637, USA} 

\author{Jiayin Gu}
\email{jiayin\_gu@fudan.edu.cn}
\affiliation{Department of Physics and Center for Field Theory and Particle Physics, Fudan University, Shanghai 200438, China}
\affiliation{Key Laboratory of Nuclear Physics and Ion-beam Application (MOE), Fudan University, Shanghai 200433, China}

\author{Zhen Liu}
\email{zliuphys@umn.edu}
\affiliation{School of Physics and Astronomy, University of Minnesota, Minneapolis, MN 55455, USA}

\author{Chi Shu}
\email{cshu20@fudan.edu.cn}
\affiliation{Department of Physics and Center for Field Theory and Particle Physics, Fudan University, Shanghai 200438, China}
\affiliation{Department of Physics, University of Chicago, Chicago, IL 60637, USA}
\affiliation{Enrico Fermi Institute, University of Chicago, Chicago, IL 60637, USA}
\affiliation{Kavli Institute for Cosmological Physics,
University of Chicago, Chicago, IL 60637, USA}

\author{Lian-Tao Wang}
\email{liantaow@uchicago.edu}
\affiliation{Department of Physics, University of Chicago, Chicago, IL 60637, USA}
\affiliation{Enrico Fermi Institute, University of Chicago, Chicago, IL 60637, USA}
\affiliation{Kavli Institute for Cosmological Physics,
University of Chicago, Chicago, IL 60637, USA}

\begin{abstract}
    Higgs couplings are essential probes for physics beyond the Standard Model (BSM) since they can be modified by new physics, such as through the Higgs portal interaction $|H|^2\mathcal{O}$. These modifications influence Higgs interactions via dimension-6 operators of the form $ \left(\partial |H|^2\right)^2$ and $|H|^6$, which are generally expected to be of comparable size. This paper discusses a phenomenon of accidental suppression, where the $|H|^6$ coupling is significantly smaller than $\left(\partial |H|^2\right)^2$. This suppression, arising from the truncation of the tree-level effective potential, lacks a clear symmetry explanation but persists in portal models. This paper aims to inspire further studies on additional instances of accidental suppression without symmetry explanations or a general framework to characterize such suppression. We also discuss constraints, at the HL-LHC and future colliders, on the Wilson coefficients of the two dimension-6 operators for various benchmark scenarios of the concrete model.
\end{abstract}

\preprint{UMN-TH-4327/24}

\maketitle

{\small 
\tableofcontents}

\section{Introduction}

Understanding the impact of physics beyond the Standard Model (BSM) at energy scales accessible to current and near-future experiments is a main goal of particle physics. Following the successful detection of the Higgs boson by the ATLAS \cite{ATLAS:2012yve} and CMS \cite{CMS:2012qbp} collaborations at the Large Hadron Collider (LHC) a decade ago, precision measurements of Higgs couplings have become increasingly important as a gateway to BSM physics. The Higgs sector is of particular interest because it can serve as a portal to the new physics, which involves particles generally neutral under the Standard Model (SM) gauge group \cite{Patt:2006fw}. Higgs physics can probe many BSM models, including dark matter and dark sector physics, through the Higgs portal. Since the proposal of a dark sector that couples primarily to the Higgs, extensive efforts have been devoted to studying benchmark models with an extended scalar sector via Higgs portal couplings \cite{Henning:2014wua, deBlas:2014mba, Gorbahn:2015gxa, Brehmer:2015rna, Carena:2013ooa, Buchalla:2016bse, Dawson:2017vgm, deBlas:2017xtg, Gori:2018pmk, Fuchs:2020cmm, Marzocca:2020jze, Cohen:2020xca, Dawson:2021jcl}. For reviews on these models, see Refs.~\cite{LHCHiggsCrossSectionWorkingGroup:2016ypw, Arcadi:2019lka}.

It is necessary to systematically quantify the impact of new physics sectors on Higgs physics. Collider probes offer leading constraints on deviations of Higgs couplings from their SM values. Numerous studies have explored how various channels can measure possible deviations at the LHC~\cite{Elias-Miro:2013mua, Pomarol:2013zra, Ellis:2014dva, Falkowski:2015fla, Campbell:2017rke, Maltoni:2017ims, Carena:2018vpt, Biekotter:2018ohn, Arganda:2018ftn, Carena:2019une, DiMicco:2019ngk, Haisch:2021hvy, Ellis:2020unq, Carena:2021onl, Ethier:2021bye, Carena:2022yvx} and at future lepton colliders \cite{Barger:2003rs, Han:2012rb, Han:2013kya, McCullough:2013rea, Craig:2014una, Craig:2015wwr, Ellis:2015sca, Ge:2016zro, Greco:2016izi, Voigt:2017vfz, Lafaye:2017kgf, Durieux:2017rsg, dEnterria:2017dac, Barklow:2017awn, Barklow:2017suo, Gu:2017ckc, Chiu:2017yrx, DiVita:2017vrr, An:2018dwb, deBlas:2019rxi, DeBlas:2019qco, deBlas:2022aow, Forslund:2022xjq, deBlas:2022ofj}.

Effective field theory (EFT) provides a robust framework for discussing the impact of high-energy physics observed at lower energy scales. These impacts are usually imprinted as higher-dimensional operators suppressed by a cutoff scale introduced by the ultraviolet (UV) physics. From the infrared (IR) EFT perspective, all higher-dimensional operators should be consistently included in the EFT as long as they are permitted by the symmetry. Conversely, from the UV perspective, certain EFT operators in the IR can be forbidden by imposing symmetries in the UV, which may be obscure in the IR. The symmetry-based suppression is well-known and is applied to construct natural solutions to the hierarchy problem via supersymmetry, compositeness, or extra dimensions (see Refs.~\cite{Martin:1997ns, Panico:2015jxa, Agrawal:2022rqd} for reviews on these topics). It is also used to investigate non-supersymmetric non-renormalization theorems \cite{Alonso:2014rga, Elias-Miro:2014eia, Cheung:2015aba}, solve the strong CP problem via continuous PQ symmetry (see Refs.~\cite{Kim:2008hd, Hook:2018dlk, Blinov:2022tfy} for overviews) or some discrete symmetries \cite{Beg:1978mt, Babu:1988mw, Nelson:1983zb, Barr:1984qx, Barr:1991qx, Bento:1991ez, Hiller:2001qg, Albaid:2015axa, Hall:2018let, deVries:2021pzl, Craig:2020bnv}, etc. In contrast, the suppression of Wilson coefficients without symmetry reasons is sometimes considered unnatural and may imply tuning of parameters in the UV theory, as enforcing the absence of higher-dimensional operators without some symmetry is generally challenging. It is, therefore, particularly interesting to explore whether one can suppress EFT operators without symmetries and if such suppression is relevant to searches for new physics.

In this paper, we present a mechanism to realize the accidental suppression of Wilson coefficients for higher-dimensional operators without a clear symmetry protecting the operator. This accidental suppression is related to the polynomial nature of the classical solution to the equation of motion. When the classical solution truncates at a certain order in the light degrees of freedom, the tree-level effective Lagrangian also truncates, suppressing higher-dimensional operators. While this peculiar truncation of the classical solution seems to require some tuning of the model parameters in the UV, we provide a concrete example of such suppression in the context of the SM extended with an additional (SM) singlet scalar. This model requires no tuning of the UV parameters to achieve accidental suppression, and the singlet can act as the portal coupling to any hidden sector not charged under the SM, which can also be a generic dark sector Higgs field charged under new dark gauge groups. The accidental suppression in the singlet extension manifests in two dimension-6 SMEFT operators, which can be probed by future colliders performing precision Higgs measurements, highlighting the synergy between the Higgs precision physics and BSM physics. This example, relevant to many Higgs-portal dark-sector models, also challenges the naïve expectation that symmetry is necessary to enforce the suppression of Wilson coefficients.

This paper is organized as follows. \Cref{sec:UnexpectedSuppression} begins with an explanation of how EFT organizes higher-dimensional operators and offers a review of the tree-level and one-loop effective Lagrangian. We then apply techniques discussed in \Cref{sec:EffLagReview} to toy models in \Cref{sec:TruncatedExamples} to demonstrate how one realizes a truncated Lagrangian to suppress Wilson coefficients. A concrete example applicable to BSM physics is presented in \Cref{sec:RealisticModel}, requiring neither tuning of UV parameters nor special symmetry to protect higher-dimensional operators in the Higgs sector. Future-collider measurements of these Wilson coefficients are discussed in \Cref{sec:ColliderPheno}. \Cref{fig:WilsonCoef} highlights the interplay between the theoretical expectation of Wilson coefficients and future experimental probes via precision Higgs measurements. Finally, we summarize our results in \Cref{sec:Conclusion}.

\section{Expected Relation Confronting Unexpected Suppression \label{sec:UnexpectedSuppression}}

Starting with a consistent UV model, one generally expects a relationship between operators due to their shared symmetries, leading to comparable Wilson coefficients. For instance, after integrating out a heavy particle $S$, its effects on the light field $H$ can be parameterized by replacing the $S$ propagator with its Taylor expansion in terms of $\sim \order{1/m_S^2}$, where $m_S$ denotes the mass of the heavy particle $S$. Consequently, the operators in the low-energy EFT are expected to respect, at least at tree level, some relations indicating their origin from the propagator of $S$, i.e.,
\begin{equation}
    \begin{multlined}
        \Lag[UV] = \frac{1}{2} S \left(-\square - m_S^2\right) S + S J(H) \\
        \to 
        \Lag[EFT] = J(H) \frac{1}{-\square - m_S^2} J(H) 
        = -\frac{1}{2} \left(\frac{J^2}{m_S^2} + \frac{J \square J}{m_S^4} + \ldots\right),
    \end{multlined}
\end{equation}
where $\square \defeq \partial_\mu \partial^\mu = -p^2$ denotes the d'Alembertian operator. This suggests that operators with derivatives, such as $J \square J / m_S^4$ in the equation above, are generally expected to generate Wilson coefficients of the same order as $J^2 / m_S^2$. One might even suspect that knowing all non-derivative operators would allow enforcing some constraints on the Wilson coefficients of operators with derivatives. After all, locality enforces the $S$ propagator to take the form $1/(p^2 - m_S^2)$, which is then expanded in a Taylor series in the EFT formalism. 

However, as we will discuss, this needs not be the case. The accidental suppression of Wilson coefficients can occur without any symmetry protection. We will first review the tree-level and one-loop effective Lagrangian. Then, we will demonstrate how this accidental suppression can occur, accompanied by two simple examples under analytic control. Finally, we will show that this mechanism is also present in a simple BSM benchmark model with no tuned parameters in \cref{sec:RealisticModel}.

\subsection{Tree-level and One-loop Effective Lagrangian \label{sec:EffLagReview}}

The effective action of a UV theory described by $\mathcal{S}_{\text{UV}}$ can be defined as 
\begin{equation}
    e^{i \mathcal{S}_{\text{EFT}}(H)} \propto \int \mathcal{D} S \; \exp( i \mathcal{S}_{\text{UV}}(H, S) ),
\end{equation}
where we integrate out the heavy field $S$ to obtain an effective action solely for the light field $H$. To find $\mathcal{S}_{\text{EFT}}$, a saddle-point approximation around the classical solution $S_c(H)$ is performed. For the non-derivative terms at the tree level, this procedure is intuitive. The classical solution is obtained by minimizing the UV potential and substituting the heavy field with its classical solution, resulting in the tree-level non-derivative effective action
\begin{equation}
    \mathcal{S}_{\text{EFT}} \supset -\int \mathrm{d}^4 x\; V(H,S_c(H)) \quad \text{with} \quad 0 = \left. \frac{\partial V(H,S)}{\partial S} \right|_{S = S_c(H)},
\end{equation}
where $V(H,S)$ denotes the non-derivative terms in the UV theory.

To generalize this computation to include derivative operators requires some care. We use the long-wavelength expansion by scaling $x \to x / \epsilon$ and $\partial_\mu \to \epsilon \partial_\mu$ to implement an order-by-order expansion in derivative operators. We assume that the UV action, effective action, and the classical solution all admit perturbative expansions in $\order{\epsilon}$:
\begin{equation}
    \mathcal{S}_{\text{UV/EFT}} = \sum_{n = 0}^{\infty} \epsilon^{2n} \mathcal{S}_{\text{UV/EFT}}^{(2n)}, \quad S_c(H) = \sum_{n = 0}^{\infty} \epsilon^{2n} S_c^{(2n)}(H).
\end{equation}
With this rescaling, we separate the action with no derivative operators $S^{(0)}$ from those with two derivative operators $S^{(2)}$. Lorentz invariance dictates that derivative operators appear in pairs for bosonic operators, justifying the expansion in even powers of $\epsilon$. By requiring the saddle-point condition to hold order-by-order in $\epsilon$
\begin{equation}
    0 = \frac{\delta \mathcal{S}_{\text{UV}}}{\delta S}\,(H, S_c) 
    = \sum_{n} \epsilon^{2n} \frac{\delta \mathcal{S}_{\text{UV}}^{(2n)}}{\delta S}\,(H, S_c^{(0)} + \epsilon^2 S_c^{(2)} + \ldots),
    \label{eqn:derivPerturbSc}
\end{equation}
we obtain the following relation between $\mathcal{S}_{\text{UV}}$ and the tree-level $\mathcal{S}_{\text{EFT}}$
\begin{equation}
    \mathcal{S}_{\text{EFT}}^{(0)} = \left.\mathcal{S}_{\text{UV}}^{(0)}\right|_{S = S_c^{(0)}}, \quad 
    \mathcal{S}_{\text{EFT}}^{(2)} = \left.\mathcal{S}_{\text{UV}}^{(2)}\right|_{S = S_c^{(0)}}, \quad 
    \mathcal{S}_{\text{EFT}}^{(4)} = -\left.\frac{1}{2} \frac{\delta \mathcal{S}_{\text{UV}}^{(2)}}{\delta S} \left[ \frac{\delta^2 \mathcal{S}_{\text{UV}}^{(0)}}{\delta S^2} \right]^{-1} \frac{\delta \mathcal{S}_{\text{UV}}^{(2)}}{\delta S}\right|_{S = S_c^{(0)}}, \ldots
\end{equation}
For our discussion, it suffices to consider up to $\order{\epsilon^2}$.

For the one-loop contribution, we follow the computation in Appendix D of Ref.~\cite{Cohen:2020xca}, which evaluates the functional trace of the one-loop integral using covariant derivative expansion \cite{Gaillard:1985uh, Cheyette:1987qz, Henning:2014wua, delAguila:2016zcb, Henning:2016lyp, Camargo-Molina:2016moz, Fuentes-Martin:2016uol, Zhang:2016pja, Brivio:2017vri, Ellis:2017jns, Jiang:2018pbd, Cohen:2020fcu}. The idea is to evaluate the operator appearing in the heavy propagator $\mathcal{O} = - \delta^2 \mathcal{S}_{\text{UV}}/\delta S^2 = \partial^2 + M^2 + U(x)$, where $M$ denotes the mass of the heavy field and $U$ is some general local operator (which may depend on the light field $H$). The effective action at loop level is treated using the background field method $S \to S_c + S_q$, where $S_c$ denotes the classical background, and $S_q$ the quantum field fluctuations. The saddle-point approximation allows us to expand the UV action as
\begin{equation}
    \begin{aligned}
    e^{i\mathcal{S}_{\text{EFT}}(H)} 
    \propto& \int \mathcal{D} S_q \; \exp(i \mathcal{S}_{\text{UV}}(S_c) + \frac{i}{2} \left.\frac{\delta^2 \mathcal{S}_{\text{UV}}}{\delta S^2}\right|_{S_c} S_q^2 + \ldots ) \\
    \approx& \exp(i\mathcal{S}_{\text{UV}}(S_c)) \det(\mathcal{O})^{-1/2} = \exp\left[i \left(\mathcal{S}_{\text{UV}}(S_c) + \frac{i}{2} \Tr\ln\mathcal{O} \right)\right]
    \end{aligned}
\end{equation}
Then, the one-loop effective action by tracing over $\ln\mathcal{O}$ becomes 
\begin{equation}
    \begin{aligned}
        \mathcal{S}_{\text{1-loop}} =& \frac{i}{2} \Tr\ln\mathcal{O} = \frac{1}{2} \int \mathrm{d}^4 x\; \frac{1}{(4\pi)^2} \Tr\left[ M^2 \left(\ln\left(\frac{\mu^2}{M^2}\right) + 1\right) U + \left(\frac{1}{2} \ln\left(\frac{\mu^2}{M^2}\right)\right) U^2 \right. \\
        & \left. - \frac{U^3}{6M^2} + \frac{(\partial U)^2}{12 M^2} \right] + \order{\frac{1}{M^4}},
    \end{aligned}
    \label{eqn:OneLoopActEff}
\end{equation}
where the trace is over indices on $U$. In our discussion, we will pay special attention to the last two terms of order $\order{1/M^2}$. These irrelevant operators, while not sensitive or marginally sensitive to the UV theory, can be seen as indicative of new physics since such terms do not exist in the Standard Model. In particular, one may enforce a renormalization condition in the IR to match the Wilson coefficients of the relevant and marginal operators in this EFT to those in the Standard Model, so the dominant effect of new physics on these operators comes from the renormalization group (RG) running, which may be slow due to the logarithmic dependence. However, the existence of higher-dimensional operators due to the finite piece in the loop correction cannot be attributed to the RG running in the Standard Model, thus signaling new physics more directly.

\subsection{Truncated Tree-level Effective Lagrangian and Examples of Accidental Suppression \label{sec:TruncatedExamples}}

This section provides two examples demonstrating the accidental suppression of non-derivative operators at the tree level. This mechanism generally arises from a truncation of the effective potential. The two examples considered here have tuned parameters and do not appear to lead to a larger symmetry, illustrating why such suppression seems accidental. In the next section, we will present a realistic model that realizes this accidental suppression without fine-tuning.

Our first example involves the following Lagrangian, 
\begin{equation}
    \mathcal{L}_{\text{UV}} = -\frac{1}{2} S \square S + \frac{f(H)}{n} S^n - \frac{\lambda_{2n}}{2n} S^{2n},
    \label{eqn:Sncase}
\end{equation}
where \(n\) is a positive integer, \(f(H)\) is a polynomial in the light degrees of freedom \(H\), \(S\) is a heavy real scalar to be integrated out, and \(\lambda_{2n}\) is the coupling constant for the \(2n\)-point interaction of \(S\). The classical solution \(S_c^{(0)}\) is%
\footnote{Note that while \(S_c^{(0)} = 0\) could be a solution, one can prohibit it by making such a solution an unstable false vacuum. For instance, when \(n = 2\), if \(f(H)\) has a positive constant piece, \textit{i.e.}, \(f(H) = c_0 + c_1 H + \ldots\) with \(c_0 > 0\), then \(\partial^2 \mathcal{L}/\partial S^2 = c_0 > 0\) indicates that the potential of \(S\) near \(S_c = 0\) has a negative curvature and is unstable.}
\begin{equation}
    0 = \frac{\delta \mathcal{S}_{\text{UV}}}{\delta S} = S^{n-1} \left[f(H) - \lambda_{2n} S^n\right] \implies S_c^{(0)} = \left[\frac{f(H)}{\lambda_{2n}}\right]^{1/n}.
\end{equation}
Substituting \(S_c^{(0)}\) into \(\mathcal{S}_{\text{UV}}\), we find that 
\begin{equation}
    \mathcal{L}_{\text{EFT}} \supset \frac{\left[f(H)\right]^2}{2n \lambda_{2n}} - \frac{1}{2} \left(\frac{f(H)}{\lambda_{2n}}\right)^{1/n} \square \left(\frac{f(H)}{\lambda_{2n}}\right)^{1/n}.
\end{equation}
As long as \(f(H)\) is a polynomial of at most \(\mathcal{O}(H^m)\), the first term is truncated at order \(H^{2m}\), resulting in the \textit{vanishing} of the tree-level Wilson coefficients for all \(\mathcal{O}(H^p)\) for \(p>2m\) in the EFT. On the other hand, the $S$ field kinetic-term-induced contribution, which is the second term in the above equation, generally involves the \(n\)-th root of \(f(H)\); thus, as long as \(n \neq 1\), the effective Lagrangian should be a power series in \(H\), generally having infinitely many terms. Hence, in this example, the \(H^p\) term is accidentally suppressed with respect to operators \(\sim \partial^2 H^{p-2}\). To be more concrete, let us consider \(n = 2\) and \(f(H) = M^2 + \mu H\). Then, the effective Lagrangian would be
\begin{equation}
    \begin{aligned}
        \mathcal{L}_{\text{EFT}} \supset& \frac{\left(M^2 + \mu H\right)^2}{4\lambda_4} - \frac{1}{2\lambda_4} \sqrt{M^2 + \mu H} \square \sqrt{M^2 + \mu H} \\
        \approx& \frac{M^4}{4\lambda_4} + \frac{M^2 \mu}{4\lambda_4} \left(2H - \frac{\square}{M^2} H\right) + \frac{\mu^2}{16 \lambda_4} \left( 4H^2 - 2H \frac{\square}{M^2} H + \frac{\square}{M^2} H^2 \right) \\
        & + \frac{\mu^3}{32 \lambda_4 M^2} \left( H^2 \frac{\square}{M^2} H + H \frac{\square}{M^2} H^2 - \frac{\square}{M^2} H^3 \right) + \mathcal{O}(M^{-4}),
    \end{aligned}
\end{equation}
where we included total derivative terms such as \(\sim \square H\) and redundant terms related by integration by parts to explicitly show the relation between the non-derivative operators and those with two derivatives. The absence of the non-derivative \(H^{\geq3}\) operator in the effective Lagrangian shows accidental suppression in this theory, and this truncation happens for any integer \(n\). However, for \(n \geq 3\), the theory would require tuning as higher-dimensional operators are generally expected to contribute to the RG running of lower-dimensional operators, and the absence of operators of the form \(S^{2n-1}, \ldots, S^{n+1}\) is generally unexpected. One might argue that a discrete \(\mathbb{Z}_n\) symmetry could be imposed on the \(S^n\) operator, suppressing all operators from \(S^{2n-1}\) to \(S^{n+1}\) by symmetry, making it technically natural. However, this is not quite the case, as illustrated in the next example.

Let us consider another example with the following Lagrangian
\begin{equation}
    \begin{multlined}
        - \mathcal{L} = \mu_h^2 |H|^2 + \frac{1}{2} \mu_s^2 S^2 + \frac{\lambda_h}{4} |H|^4 + \frac{\lambda_m}{2} |H|^2 S^2 + \frac{\lambda_s}{4} S^4 \\
        + \frac{\lambda_{s6}}{6} S^6 + \frac{\lambda_{h6}}{36} |H|^6 + \frac{\lambda_{s2h4}}{2} S^2 |H|^4 + \frac{\lambda_{s4h2}}{4} S^4 |H|^2,
    \end{multlined}
\end{equation}
where \(H\) is a light degree of freedom with a \(\mathrm{U}(1)\) symmetry and \(S\) is a heavy real scalar with a \(\mathbb{Z}_2\) symmetry \(S \to -S\) to be integrated out. The classical solution here is (for the solution with \(\langle S \rangle \neq 0\))
\begin{equation}
    \begin{multlined}
        0 = \frac{\partial \mathcal{L}}{\partial S} = - S \left[\lambda_{s6} S^4 + \left(\lambda_s + \lambda_{s4h2} |H|^2\right) S^2 + \left(\mu_s^2 + \lambda_m |H|^2 + \lambda_{s2h4} |H|^4\right)\right] \\
        \implies {S_c^{(0)}}^2 = -\frac{\lambda_s + \lambda_{s4h2} |H|^2}{2\lambda_{s6}} \pm \frac{\sqrt{\Delta}}{2\lambda_{s6}},
    \end{multlined}
\end{equation}
where 
\begin{equation}
    \Delta \defeq \left(\lambda_{s4h2}^2 - 4 \lambda_{s6} \lambda_{s2h4}\right) |H|^4 + \left(2 \lambda_s \lambda_{s4h2} - 4 \lambda_{s6} \lambda_m\right) |H|^2 + \left(\lambda_s^2 - 4 \lambda_{s6} \mu_s^2\right). \label{eq:De}
\end{equation}
Hence, the effective Lagrangian will generally include non-derivative terms such as 
\begin{equation}
    \mathcal{L} \supset c_1 \Delta^{3/2} + c_2 \left(\lambda_s + \lambda_{s4h2} |H|^2\right)^3 + c_3 \Delta^{1/2} + \ldots.
\end{equation}
Upon Taylor expanding in powers of \(|H|^2\), one generally expects \(\Delta^{3/2}\) and \(\Delta^{1/2}\) to give rise to an infinite series of \(|H|^{2n}\) terms, and it is natural to expect that \(\sim |H|^{2n}\) and \(\sim \partial^2 H^{2(n-1)}\) have comparable Wilson coefficients. However, if \(\Delta\) happens to be a complete square of a polynomial in \(|H|^2\), then accidental suppression of \(|H|^{2n}\) operators occurs. The condition for \(\Delta\) to be a perfect square is 
\begin{equation}
    \left(\lambda_s \lambda_{s4h2} - 2\lambda_{s6}\lambda_m\right)^2 - \left(\lambda_s^2 - 4\lambda_{s6}\mu_s^2\right)\left(\lambda_{s4h2}^2 - 4\lambda_{s6}\lambda_{s2h4}\right) = 0
    \label{eqn:s6SumRule}
\end{equation}
so that 
\begin{equation}
    \Delta = \left(\lambda_{s4h2}^2 - 4 \lambda_{s6}\lambda_{s2h4}\right) \left[ |H|^2 + \frac{2\lambda_s \lambda_{s4h2} - 4 \lambda_m \lambda_{s6}}{2\left(\lambda_{s4h2}^2 - 4 \lambda_{s6} \lambda_{s2h4}\right)} \right]^2.
\end{equation}
In this case, the effective Lagrangian takes the form of a finite polynomial:
\begin{equation}
    \mathcal{L}_{\text{EFT}} \supset c_1 |H|^6 + \ldots.
\end{equation}
Since the sum rule presented in \cref{eqn:s6SumRule} generally requires coefficients in the UV theory across different dimensions to conspire to such an accidental cancellation, we argue that it would require considerable tuning to achieve such a truncated effective Lagrangian. More importantly, we see that the \(\mathbb{Z}_2\) symmetry on \(S\) does not enforce this truncation; rather, it is the polynomial nature of the classical solution \(S_c\)~%
\footnote{Equivalently, it is the polynomial nature of the classical solution of \(S_c^2\), for this constructed example, or any model that has a (explicit or spontaneously broken) $S\rightarrow -S$ \(\mathbb{Z}_2\) symmetry.} 
that leads to the suppression of higher-dimensional non-derivative operators. At this point, accidental suppression appears to be related only to finely tuned parameter space in the UV theory and is not applicable to a broader class of models. However, there are other simpler cases. For instance, if one sets $\lambda_{s6}$ and $\lambda_{s4h2}$ to zero, 
the solution would be truncated as well. This can be achieved by restricting the Lagrangian to be renormalizable (assuming there is a large separation between the mass of $S$ and the scale for the potential additional new physics) without any tuning of the parameters. We will study this case in detail in the next section in the context of a realistic singlet extension to the Standard Model.

\section{Simplest Benchmark Model: SM Gauge Singlet Extension \label{sec:RealisticModel}}

We now investigate one of the simplest BSM benchmark models: the singlet extension to the Standard Model, where a neutral real scalar \(S\) is introduced alongside the Standard Model particles. The most general renormalizable Lagrangian describing the interaction between the new scalar \(S\) and the Standard Model is given by its interactions with the Higgs field,
\begin{equation}
    - \mathcal{L}_{\text{UV}} \supset -\mu_{H}^{2} |H|^{2} + \lambda_{H}|H|^{4} \pm \frac{1}{2} \mu_{s}^{2} S^{2} + \frac{1}{3!} A_{s} \mu_s S^{3} + \frac{1}{4!} \lambda_{s} S^{4} + A_{m} \mu_s S|H|^{2} + \frac{1}{2} \lambda_{m} S^{2} |H|^{2},
    \label{eqn:SingletGeneral}
\end{equation}
where \(\mu_H\) and \(\mu_S\) are dimensionful couplings, while \(\lambda_H\), \(\lambda_s\), \(\lambda_m\), \(A_s\), and \(A_m\) are dimensionless couplings in the model. 

We will compare three benchmark models based on the \(\mathbb{Z}_2\) symmetry of the $S$ field, $S\rightarrow -S$: (1) exact \(\mathbb{Z}_2\) model, (2) spontaneously broken \(\mathbb{Z}_2\) model, and (3) explicitly broken \(\mathbb{Z}_2\) model. For both the exact \(\mathbb{Z}_2\) and the spontaneously broken \(\mathbb{Z}_2\) models, the parameters \(A_s = A_m = 0\) vanish due to symmetry. For the explicitly broken \(\mathbb{Z}_2\) model, all terms in \cref{eqn:SingletGeneral} are allowed. The distinction between the exact \(\mathbb{Z}_2\) model and the spontaneously broken \(\mathbb{Z}_2\) model lies in the sign of the \(\mu_S^2\) parameter: the \(\mathbb{Z}_2\) symmetry is exact when \(S\) has a positive mass, while \(\mathbb{Z}_2\) is spontaneously broken when \(S\) has a negative mass. We will demonstrate that suppression of \(|H|^{2n}\) occurs for the first two of these benchmarks. For the spontaneously broken \(\mathbb{Z}_2\) model, however, this suppression is accidental due to the truncation discussed in the previous section.

First, we examine the exact \(\mathbb{Z}_2\) model with the UV Lagrangian,
\begin{equation}
    \mathcal{L}_{\text{UV, exact}} \supset \mu_H^2 |H|^2 - \lambda_H |H|^4 - \frac{1}{2} \mu_s^2 S^2 - \frac{1}{4!} \lambda_s S^4 - \frac{1}{2} \lambda_m |H|^2 S^2.
\end{equation}
After evaluating the classical solution \(S_c^{(0)}\) in the long-wavelength limit, we find \(S_c^{(0)} = 0\). Thus, there is no tree-level contribution to the effective Lagrangian in the IR. At the one-loop level, the operator to be traced over is
\begin{equation}
    \mathcal{O} = - \left. \frac{\delta^2 \mathcal{S}_{\text{UV}}}{\delta S^2} \right|_{S = S_c} = \square + \mu_s^2 + \lambda_m |H|^2.
\end{equation}
Using \cref{eqn:OneLoopActEff}, we obtain the effective Lagrangian of the exact \(\mathbb{Z}_2\) model at the leading non-vanishing order, which is at the one-loop level
\begin{equation}
    \begin{multlined}
        \mathcal{L}_{\text{EFT, exact}} = \frac{1}{(4\pi)^2} \left[ \frac{\lambda_m \mu_s^2}{2} |H|^2 \left(\ln\left(\frac{\mu^2}{\mu_s^2}\right) + 1\right) + \frac{\lambda_m^2}{4} |H|^4 \ln\left(\frac{\mu^2}{\mu_s^2}\right) \right. \\
        \left. - \frac{\lambda_m^3}{12} \frac{|H|^6}{\mu_s^2} + \frac{\lambda_m^2}{12} \frac{(\partial |H|^2)^2}{2 \mu_s^2} + \mathcal{O}(\mu_s^{-4}) \right].
    \end{multlined}
    \label{eqn:ExactZ2ActEff}
\end{equation}
While the \(|H|^{2n}\) operators at the tree level are suppressed, this suppression is trivial and expected compared to the tree-level \(\partial^2 |H|^{2n-2}\) operators, which also vanish. At the one-loop level, non-derivative operators do not vanish following a computation similar to that by Coleman and Weinberg \cite{Coleman:1973jx}, and they have a comparable scale to their counterparts with two derivative operators. Hence, from the EFT perspective, the Wilson coefficients are only loop-suppressed and not accidentally suppressed.

In \textit{sharp} contrast, the spontaneously broken \(\mathbb{Z}_2\) model exhibits the accidental suppression. In this case, the Lagrangian becomes
\begin{equation}
    \mathcal{L}_{\text{UV, SSB}} \supset \mu_H^2 |H|^2 - \lambda_H |H|^4 + \frac{1}{2} \mu_s^2 S^2 - \frac{1}{4!} \lambda_s S^4 - \frac{1}{2} \lambda_m |H|^2 S^2,
\end{equation}
so the classical solution is
\begin{equation}
    {S_c^{(0)}}^2 = \frac{6\mu_s^2 - 6\lambda_m |H|^2}{\lambda_s},
\end{equation}
which results in a tree-level effective Lagrangian
\begin{equation}
    \begin{multlined}
        \mathcal{L}_{\text{EFT, SSB}} \supset \left(\mu_H^2 - \frac{3\lambda_m \mu_s^2}{\lambda_s}\right) |H|^2 - \left(\lambda_H - \frac{3\lambda_m^2}{2\lambda_s}\right) |H|^4 - \frac{3}{\lambda_s} \sqrt{\mu_s^2 - \lambda_m |H|^2} \square \sqrt{\mu_s^2 - \lambda_m |H|^2} \\
        \approx \left(\mu_H^2 - \frac{3\lambda_m \mu_s^2}{\lambda_s}\right) |H|^2 - \left(\lambda_H - \frac{3\lambda_m^2}{4\lambda_s}\right) |H|^4 + \frac{3\lambda_m^2}{2\lambda_s} \frac{(\partial |H|^2)^2}{2\mu_s^2} + \frac{3\lambda_m^3}{2\lambda_s} \frac{|H|^2 (\partial |H|^2)^2}{2\mu_s^4} + \mathcal{O}(\mu_s^{-6}).
    \end{multlined}
\end{equation}
Similar to the first example in \cref{sec:TruncatedExamples}, the square root leads to generally non-vanishing \(\sim \partial^2 |H|^{2n}\) operators, while the non-derivative operator is truncated at \(\sim \mathcal{O}(|H|^4)\). Thus, \(|H|^{2n}\) for \(n \geq 3\) are accidentally suppressed in the EFT. In this case, the leading contribution to \(\sim |H|^6\), starting at the one-loop level, reads
\begin{equation}
    \mathcal{L}_{\text{EFT, SSB}} \supset -\frac{1}{(4\pi)^2} \frac{\lambda_m^3}{24} \frac{|H|^6}{\mu_s^2}.
    \label{eqn:SSBLoop}
\end{equation}
We see that the \(|H|^6\) operator is unusually small compared to the \(\sim \partial^2 |H|^4\) term
\begin{equation}
    \mathcal{L}_{\text{EFT, SSB}} \supset \frac{3\lambda_m^2}{2\lambda_s} \frac{(\partial |H|^2)^2}{2\mu_s^2}
    \label{eqn:SSBTree}
\end{equation}
from the IR perspective. 

Although one might argue that the \(\mathbb{Z}_2\) symmetry, albeit spontaneously broken, enforces some residual symmetry that suppresses the tree-level operator, we remind readers that \(\mathbb{Z}_2\) cannot enforce such suppression, as seen in the second example in \cref{sec:TruncatedExamples}. In fact, requiring renormalizable \(\mathbb{Z}_2\) symmetric \(S\) accidentally satisfies the peculiar condition shown in \cref{eqn:s6SumRule}. Were we to consider a \(\mathbb{Z}_2\)-symmetric theory with irrelevant operators due to some other UV completion or a \(\mathbb{Z}_2\)-symmetric theory in some 3D QFT, this truncation at the tree level cannot occur unless parameters are tuned, demonstrating that the theory has no symmetry suppressing \(|H|^{2n}\) in general. We will discuss in \cref{sec:ColliderPheno} how to leverage the accidental suppression to probe the UV physics using dimension-6 SMEFT operators relevant to the Higgs physics at future colliders. We also emphasize that the spontaneously broken \(\mathbb{Z}_2\) model can be immediately applied to any hidden-sector messenger field or dark Higgs field that can be charged under non-SM gauge groups, \(X = S\) which couples to the Standard Model through the Higgs portal, and the accidental suppression also occurs for interactions between the Higgs and the messenger (or the radial part of the messenger).

For completeness, we now discuss the explicitly broken \(\mathbb{Z}_2\) model. The model has the following Lagrangian,
\begin{equation}
    \mathcal{L}_{\text{UV, explicit}} \supset \mathcal{L}_{\text{UV, SSB}} - \frac{1}{3!} A_s \mu_s S^3 - A_m \mu_s S |H|^2,
\end{equation}
which includes both the spontaneous symmetry breaking of the \(\mathbb{Z}_2\) symmetry and explicit \(\mathbb{Z}_2\) breaking effects. The explicit breaking of \(\mathbb{Z}_2\) selects a unique vacuum \(\langle S \rangle\). Without loss of generality, we require \(A_s > 0\) so that \(\langle S \rangle < 0\) is the true vacuum. After shifting to \(\langle S \rangle\), the remaining \(S\) has its mass squared of
\begin{equation}
    m_s^2 = \left[2 + \frac{A_s}{4\lambda_s}\left(3A_s + \sqrt{9 A_s^2 + 24 \lambda_s}\right)\right] \mu_s^2,
    \label{eqn:msSquared}
\end{equation}
which can be required to be larger than the Higgs mass so that \(S\) can be sensibly integrated out. After enforcing \(\mathcal{S}_{\text{EFT}}(H) \approx \mathcal{S}_{\text{UV}}(S_c(H))\), we obtain
\begin{equation}
    \begin{gathered}
        \mathcal{L}_{\text{EFT, explicit}} \supset -\tilde{c}_6 \frac{|H|^6}{\mu_s^2} + \tilde{c}_H \frac{(\partial |H|^2)^2}{2\mu_s^2}, \\
        \tilde{c}_H = \frac{4\left( \lambda_m \left(3A_s + \sqrt{9 A_s^2 + 24\lambda_s}\right) - 2\lambda_s A_m\right)^2}{\left( A_s \left(3A_s + \sqrt{9 A_s^2 + 24\lambda_s}\right) + 8\lambda_s\right)^2}, \\
        \tilde{c}_6 = \left(\frac{A_s^2 A_m + 4\lambda_s A_m - 2 A_s \lambda_m}{4\sqrt{9A_s^2 + 24\lambda_s}} - \frac{A_s A_m}{12}\right) \tilde{c}_H.
    \end{gathered}
    \label{eqn:ExplicitZ2ActEff}
\end{equation}
Since \(S_c(H)\) generally involves cubic roots of polynomials of \(|H|^2\), the Taylor expansion near \(H = 0\) is not truncated for both non-derivative operators and those with two derivatives, and \(|H|^{2n}\) operators are generally reintroduced to the EFT at the tree level.

\section{Collider Phenomenology Prospects \label{sec:ColliderPheno}}

In the previous section, we considered three benchmarks for the singlet extension to the Standard Model. The interesting benchmark is the spontaneously broken \(\mathbb{Z}_2\) model, where the accidental suppression of \(|H|^{2n}\) operators for \(n \geq 3\) occurs without tuning the model parameters. This accidental suppression also leads to an unexpected hierarchy between the Wilson coefficients for \(|H|^6\) and \(\left(\partial |H|^2\right)^2\) operators. Without directly accessing the UV physics at a collider, measuring nonzero Wilson coefficients may not only hint at new physics, such as the existence of the scalar singlet, but also shed light on the possible UV symmetry pattern if both dimension-6 operators can be measured. In this section, we expand on this idea and illustrate how future colliders can provide strong probes into the UV physics through measurements of Higgs interactions.

\begin{table}
\centering
    \begin{tabular}{c | *{4}{c}}
    \hline \hline
      & \multicolumn{2}{c}{$c_H$} & \multicolumn{2}{c}{$c_6$} \\
    \hline
    Exact $\mathbb{Z}_2$ & 1-loop & [\cref{eqn:ExactZ2ActEff}] & 1-loop & [\cref{eqn:ExactZ2ActEff}] \\
    SSB $\mathbb{Z}_2$ & tree & [\cref{eqn:SSBTree}] & 1-loop & [\cref{eqn:SSBLoop}] \\
    $\cancel{\mathbb{Z}_2}$ & tree & [\cref{eqn:ExplicitZ2ActEff}] & tree & [\cref{eqn:ExplicitZ2ActEff}] \\
    \hline \hline
    \end{tabular}
\caption{Pattern of Wilson coefficients in various benchmark scenarios of the singlet extension.}
\label{tab:WilsonPattern}
\end{table}

Continuing our discussion on the singlet extension, we assume that \(S\) has a larger mass than the Higgs so that the EFT treatment of integrating out \(S\) is valid. Up to dimension-6 operators, we consider the following effective Lagrangian,
\begin{equation}
    \mathcal{L}_{\text{SMEFT}} \supset |D H|^2 + \mu^2 |H|^2 - \lambda |H|^4 + c_H \mathcal{O}_H + c_6 \mathcal{O}_6, \quad 
    \mathcal{O}_H = \frac{\left(\partial |H|^2\right)^2}{2v_\text{SM}^2}, \quad \mathcal{O}_6 = -\frac{|H|^6}{v_\text{SM}^2},
\end{equation}
which encodes all impacts on the Standard Model from integrating out \(S\) at the dimension-6 level. The three benchmark scenarios produce the Wilson coefficients, \(c_H\) and \(c_6\), at different orders in perturbation theory, as shown in \cref{tab:WilsonPattern}. Notably, while the exact \(\mathbb{Z}_2\) scenario and explicitly broken \(\mathbb{Z}_2\) scenario yield \(c_H \sim c_6\), \(c_H\) is parametrically larger than \(c_6\) in the spontaneously broken \(\mathbb{Z}_2\) model due to the accidental suppression.

In each of the three benchmark scenarios, we adjust \(\lambda_H\) and \(\mu_H\) such that all relevant and marginal operators in the EFT have the same Wilson coefficients as those in the Standard Model. We also impose appropriate renormalizable conditions around the electroweak scale so that the logarithmic RG running in all the renormalizable operators is suppressed. The new physics contribution mainly comes from \(c_6\) and \(c_H\) as a function of the UV parameters. To understand the pattern of \(c_6\) and \(c_H\) better, we take random samples of the dimensionless UV parameters for the three benchmarks under a few constraints. 

\begin{figure}
    \centering
    \includegraphics[width=\linewidth]{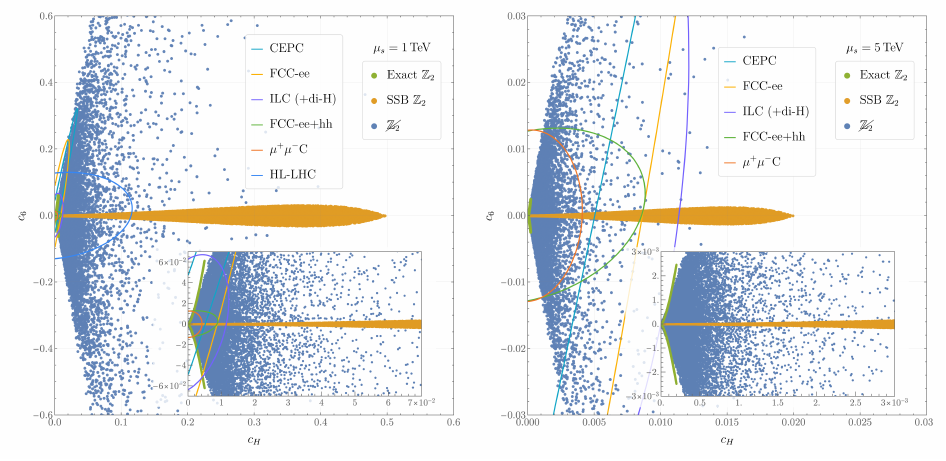}
    \caption{Wilson coefficients \((c_H, c_6)\) of dimension-6 operators generated by randomly sampled dimensionless parameters in the UV theory for three benchmark scenarios considered in \cref{sec:RealisticModel}, subject to constraints shown in \cref{eqn:ParamConstraints}. Elliptical contours (colored online) show 2$\sigma$ exclusion limits on \(c_H\) and \(c_6\) from future colliders. According to Ref.~\cite{Cepeda:2019klc}, the HL-LHC can constrain triple Higgs coupling to 50\% precision and $hZZ$ coupling coupling to 2.9\% precision at 1$\sigma$ level. The corresponding run scenarios of the future lepton colliders are listed in \Cref{tab:FutureColliderBenchmarks}. Note that the ILC constraint appears better than those from CEPC and FCC-ee due to its di-Higgs channel at \(\sqrt{s} = 500\;\mathrm{GeV}\). A zoomed-in view of the data points is available in the lower left corner of each panel to show the one-loop level Wilson coefficients for the exact \(\mathbb{Z}_2\) model. \textbf{Left:} the UV scale is set to \(\mu_s = 1\;\mathrm{TeV}\). \textbf{Right:} the UV scale is set to \(\mu_s = 5\;\mathrm{TeV}\).}  
    \label{fig:WilsonCoef}
\end{figure}

First, the potential of the singlet $S$ must be bounded from below, translating to \(\lambda_s > 0\). Second, the squared mass of \(S\) around its VEV, shown in \cref{eqn:msSquared}, must be larger than that of the Higgs boson. Here, we require \(m_s > 3m_h\) as a benchmark. Then, upon choosing the vacuum for \(S\) in the explicitly broken \(\mathbb{Z}_2\) scenario, we assume \(A_s > 0\). Lastly, to work within the perturbative regime, we enforce the tree-level \(s\)-wave unitarity so that all dimensionless UV parameters have appropriate upper bounds. More details on unitarity bounds can be found in \cref{app:UnitarityBounds}. These constraints are summarized as follows,
\begin{equation}
    \begin{gathered}
        \lambda(\lambda_H, \lambda_s, \ldots, A_m) = \lambda_\text{SM} = \frac{m_h^2}{2v_\text{SM}^2}, \quad 
        \mu^2(\lambda_s, \ldots, A_m, \mu_s, \mu_H) = \frac{m_h^2}{2}, \\ 
        \lambda_s > 0, \quad m_s(\lambda_s, \ldots, A_m, \mu_s) \geq 3m_h, \quad A_s \geq 0, \quad
        \lambda_s \leq 16\pi, \quad \lambda_H \leq \frac{8\pi}{3}, \\ 
        \lambda_m \leq 8\pi, \quad
        6 \lambda_H + \lambda_s + \sqrt{\left(6 \lambda_m - \lambda_s\right)^2 + 4 \lambda_m^2} \leq 32\pi.
    \end{gathered}
    \label{eqn:ParamConstraints}
\end{equation}
Under these constraints, we sample uniformly over \(\{\lambda_s, \lambda_m, A_s, A_m\}\) while, for simplicity, setting a few benchmark values for \(\mu_s\). The resulting \(c_H\) and \(c_6\) as a function of the UV parameters are shown as scattered points in \cref{fig:WilsonCoef}.%
\footnote{
Note that $\mu_s^2 < 0$ is not required when presenting the distribution of Wilson coefficients from a UV model with an explicitly broken $\Z[2]$. 
}%
\footnote{We also note here that, in principle, through extreme tuning, the blue scattered points should cover the orange region, as the post-SSB Lagrangian parameters represent a subset of explicit $\Z[2]$ breaking theory. However, the needed tuning without UV origins, such as SSB, is not generic; hence, we see a very low density of blue points overlapping with the orange regions.}
Currently, ATLAS \cite{ATLAS:2023qzf, ATLAS:2023gzn} and CMS \cite{CMS:2021nnc, CMS:2022cpr} provide \(\sim \mathcal{O}(1)\) exclusion at 2$\sigma$ level on these two Wilson coefficients. If future machines, such as FCC-ee, CEPC, ILC, FCC-hh, and a muon collider, are realized, the measurement on Higgs self-coupling can be significantly improved. We follow previous studies providing strategies \cite{Barklow:2017awn, Degrassi:2016wml, Gu:2017ckc, deBlas:2019rxi, Contino:2016spe, Banerjee:2018yxy, Li:2024joa, Han:2020pif} and benchmarks \cite{Bernardi:2022hny, CEPCPhysicsStudyGroup:2022uwl, DiVita:2017vrr, Barklow:2015tja, Contino:2016spe, AlAli:2021let, Black:2022cth} for detecting \(c_6\) and \(c_H\). The 2$\sigma$ exclusion limits on \(c_H\) and \(c_6\) from these future colliders are shown as ellipses in \cref{fig:WilsonCoef}. More details on constraints from future colliders are discussed in \cref{app:FutureColliderConstraints}. 

\Cref{fig:WilsonCoef} shows the features discussed in \cref{sec:RealisticModel}. When the \(\mathbb{Z}_2\) symmetry of the singlet is exact, both \(c_H\) and \(c_6\) vanish at the tree level, and loop-induced \(c_H\) and \(c_6\) (green dots in \cref{fig:WilsonCoef}) are highly suppressed. In general, even if \(\mu_s = 1\;\mathrm{TeV}\), measuring \(c_H\) and \(c_6\) for the exact \(\mathbb{Z}_2\) model is challenging in future lepton colliders, unless a high-energy collider such as FCC-hh, \(500\;\mathrm{GeV}\) ILC, or high-energy muon collider offers a synergy to measure the multi-Higgs production channel. However, when the quartic couplings are strong, both CEPC and FCC-ee can probe some combination of \(c_6\) and \(c_H\) from associated \(hZ\) production, as shown in the lower-right corner of the left panel of \cref{fig:WilsonCoef}. For the other two benchmark scenarios, Wilson coefficients are produced at the tree level, and future colliders can generally provide decent measurements of them. 

In particular, the spontaneously broken \(\mathbb{Z}_2\) model has the accidental suppression of \(c_6\), and the Wilson coefficients for this scenario (yellow dots in \cref{fig:WilsonCoef}) are predominantly distributed near \(c_6=0\). While this scenario can be considered a special case of the explicitly broken \(\mathbb{Z}_2\) model with \(A_s = A_m = 0\), it is unlikely that \(c_6\) is suppressed for the most general model (blue dots in \cref{fig:WilsonCoef}). Thus, if one focuses on the singlet extension to the Standard Model, measuring nonzero \(c_6\) and \(c_H\) not only signals the existence of new physics but also hints at the specific symmetry-breaking pattern in the UV. In general, the ratio \(c_6 / c_H\) is roughly \(\mathcal{O}(0.1 - 1)\) for small UV couplings \(\{\lambda_s, \lambda_m, A_s, A_m\}\) and at most \(\mathcal{O}(10)\), as shown in \cref{eqn:ExplicitZ2ActEff,eqn:ExactZ2ActEff}. However, when the \(\mathbb{Z}_2\) symmetry of \(S\) is spontaneously broken, \(c_6\) is loop-suppressed relative to \(c_H\), resulting in a typical value of \(c_6/c_H \sim \mathcal{O}(10^{-3} - 10^{-2})\). Therefore, when \(c_6/c_H\) is unusually small, one may argue that either the singlet has a (at least approximate) spontaneously broken \(\mathbb{Z}_2\) symmetry,\footnote{As we emphasized throughout this paper, this accidental suppression is not directly a result of symmetries in general.} or the UV model has very tuned parameters. Indeed, one can obtain a vanishing \(c_6\) in the explicitly broken \(\mathbb{Z}_2\) scenario by either taking the \(\mathbb{Z}_2\) limit or tuning one of the UV parameters, e.g., \(\lambda_m \to 2 \lambda_s A_m / A_s + A_m \left(3 A_s - \sqrt{9 A_s^2 + 24 \lambda_s}\right) / 6\). However, this tuning is unlikely if we treat all UV parameters \(\{\lambda_s, \lambda_m, A_s, A_m\}\) on equal footing. 

It is also worth mentioning that the accidental truncation of Wilson coefficients can complement the resonance search of a singlet scalar. Many strategies for searching for heavy singlet scalars at various future hadron and lepton colliders via direct production have been proposed \cite{Buttazzo:2015bka, Buttazzo:2018qqp, Carena:2018vpt, Adhikari:2020vqo, Davoudiasl:2021syn, AlAli:2021let}. Yet, even if one observes such a resonance, this heavy particle could be the radial mode \(r_s\) of some broken singlet \(S\). Both the SSB \(\mathbb{Z}_2\) scenario and the explicitly broken \(\mathbb{Z}_2\) scenario permit a coupling of the form \(\sim r_s h^2\). Hence, it is generally challenging to distinguish the two scenarios from resonance searches alone. Measuring and comparing their Wilson coefficients, on the other hand, provides more insights into the potential symmetry-breaking pattern.

\section{Conclusion and Discussion \label{sec:Conclusion}}

In this paper, we explored the accidental suppression of Wilson coefficients for non-derivative operators. Generally, one expects the \(\sim \partial^2 H^{p-2}\) operator to have comparable Wilson coefficients to those of \(\sim H^p\), as they stem from the same formal expansion of the propagator of the heavy particle being integrated out. The suppression of \(\sim H^p\) is typically attributed to symmetry considerations. However, we proposed the possibility of suppressing non-derivative operators due to the polynomial nature of the classical solution for the heavy field, independent of its symmetry. If the classical solution of the heavy field is polynomial in the light field, the tree-level effective Lagrangian is truncated at a particular order in powers of the light field, suppressing all higher-point self-interactions.

We provided two simple examples to illustrate this truncation and show the generic tuning of the UV parameters needed in general cases. We then introduced a simple model, the singlet extension to the Standard Model, that also exhibits accidental suppression. Three benchmark scenarios of the singlet extension were considered, and when a spontaneously broken \(\mathbb{Z}_2\) symmetry is enforced on the singlet, the accidental suppression naturally appears without tuning the UV parameters. It is noteworthy that, counterintuitively, the \(\mathbb{Z}_2\) symmetry cannot be responsible for the suppression, as demonstrated by comparing the second toy example from \cref{sec:TruncatedExamples} with the singlet extension. Both \(\mathbb{Z}_2\) symmetry and the termination of operators at the renormalizable level for scalars in 4D at quartic order conspire to this accidental truncation, making it independent of symmetry properties.

By leveraging the different patterns of Wilson coefficients in various benchmark scenarios, we proposed the possibility of distinguishing UV symmetry-breaking patterns by measuring the Wilson coefficients of these higher-dimensional Higgs interactions without directly accessing the UV. This highlights the impact of electroweak and Higgs precision measurements on BSM physics at future colliders.

It is intriguing that no symmetry in the Lagrangian is identified to suppress a higher-dimensional operator in the EFT, as illustrated in this paper. While the polynomial nature of the classical solution serves as a concrete but mathematical explanation for the truncation, it raises questions about whether this phenomenon is merely a mathematical coincidence with no physical explanation. Perhaps, some hidden symmetry distinguishes the \(\sim \partial^2 H^{p-2}\) operators from \(\sim H^{p}\) at the tree level. If so, we hope this paper serves as a call to seek a simple symmetry-based explanation for such accidental suppression. On the other hand, the examples we study here are to find polynomial solutions of the heavy field's equation of motion. One can use the Galois group to understand and analyze the factorizations and existence of polynomial solutions, which may help reveal the underlying physics. How the Galois group can be used and is related to the UV Lagrangian is yet to be explored.

From a model-building perspective, understanding how accidental suppression can be utilized in a broader context is interesting. Specifically, without tuning the UV parameters, are there other examples of this accidental suppression? How relevant are these examples to physics beyond the Standard Model? Is there a more general procedure to construct a model with accidental suppression? As it turns out, another simple example that leads to accidental suppression without tuning is the addition of a sterile neutrino to the Standard Model with all renormalizable interactions. Integrating out the heavy sterile neutrino yields a truncated tree-level Lagrangian, including the dimension-5 Weinberg operator. This suppression is a consequence of a trivial truncation, as the Lagrangian contains only \(\mathcal{L} \sim J S + J^\dagger S^\dagger + S^\dagger (i \cancel{\partial} + m_S) S\), and the effective potential can be derived similarly to the first toy example in \cref{sec:TruncatedExamples}.

Interestingly, both the neutrino portal and the Higgs portal exhibit accidental suppression of Wilson coefficients. However, it remains unclear how to write a general Lagrangian that guarantees truncation in its low-energy EFT. We hope this paper serves as an invitation for further studies on accidental suppression. If accidental suppression is not accidental and has a symmetry argument, understanding how such a symmetry enforces the polynomial nature of the classical solution and its model-building implications would be valuable. If accidental suppression is genuinely accidental, identifying more examples and examining whether these examples challenge the notion of technical naturalness would be exciting. If technical naturalness guides our exploration of new physics, could we miss new physics if they are accidentally suppressed?

\acknowledgments

JG and CS are supported by the National Natural Science Foundation of China (NSFC) under grant No. 12035008 and No. 12375091. ZL is supported in part by U.S. Department of Energy Grant No. DESC0011842. YB and LTW are supported by the Department of Energy grant DE-SC0013642. We provide supplemental information for this work, including the plots and scan data, on \href{https://github.com/ZhenLiuPhys/HiggsTruncation}{GitHub} under CC-BY-NC-SA.

\appendix

\section{Constraining $c_6$ and $c_H$ from Higgs Measurements at Future Colliders \label{app:FutureColliderConstraints}}

Given the following effective Lagrangian
\begin{equation}
    \mathcal{L}_{\text{SMEFT}} \supset |D H|^2 + \mu^2 |H|^2 - \lambda |H|^4 + \frac{c_H}{2v_\text{SM}^2} \left(\partial |H|^2\right)^2 - \frac{c_6}{v_\text{SM}^2} |H|^6,
\end{equation}
one can quantify the impact of the higher-dimensional operators through Higgs measurements. Assuming that \(c_6\) and \(c_H\) are small, since they usually carry a suppression factor \(\sim v_\text{SM}^2 / \Lambda_\text{UV}^2\), and with the freedom to choose the UV parameters \(\mu_H^2\) and \(\lambda_H\), we set \(\mu^2(\mu_H^2)\) and \(\lambda(\lambda_H)\) to match their expected Standard Model values.

Parameterizing the Higgs doublet in the unitarity gauge by \(H = (v_\text{SM} + h) / \sqrt{2}\), where \(h\) denotes the Higgs boson, the presence of \(|H|\) alters the Higgs potential and the Higgs VEV. The \(\left(\partial |H|^2\right)^2\) operator shifts the kinetic term of the Higgs field, containing terms like \(\sim (\partial h)^2\) and \(\sim h (\partial h)^2\), introducing an energy dependence to the trilinear Higgs interaction. To reproduce the Standard Model, we fix the electroweak VEV \(v_\text{SM} = 246 \; \text{GeV}\) and the Higgs mass \(m_h = 125 \; \text{GeV}\) according to their experimental values, which uniquely sets the quartic Higgs coupling in the Standard Model to \(\lambda_\text{SM} = m_h^2/(2v_\text{SM}^2)\).

To achieve this, the following renormalization of the coefficients \(\mu^2\) and \(\lambda\) and redefinition of the Higgs boson \(h\) are required
\begin{equation}
    \begin{gathered}
        \lambda \to \lambda_\text{SM} \left(1-\frac{3}{2} \frac{c_6}{\lambda_\text{SM}} + c_H\right), \quad
        \mu^2 \to \mu_\text{SM}^2 \left(1-\frac{3}{4}\frac{c_6}{\lambda_\text{SM}} + c_H\right), \\
        h \to \left(1 - \frac{c_H}{2} - \frac{c_H}{2}\frac{h}{v_\text{SM}} - \frac{c_H}{6} \frac{h^2}{v_\text{SM}^2}\right) h.
    \end{gathered}
\end{equation}
After implementing the renormalization and field redefinition, we find
\begin{equation}
    \mathcal{L}_{\text{SMEFT}} \supset \frac{1}{2} \left(\partial h\right)^2 - \frac{1}{2} m_h^2 h^2 - \sqrt{\lambda_\text{SM}} \mu_\text{SM} \left(1 + \frac{c_6}{\lambda_\text{SM}} - \frac{3}{2} c_H\right) h^3 + \mathcal{O}(h^4).
\end{equation}
Thus, the trilinear Higgs self-coupling is altered by the presence of \(c_6\) and \(c_H\). Additionally, due to the field redefinition of \(h\), its couplings to other SM particles receive a universal modification from \(c_H\). For instance, the \(hZZ\) coupling will be affected as:
\begin{equation}
    \mathcal{L}_{\text{SMEFT}} \supset \frac{m_Z^2}{v_\text{SM}} \left(1 - \frac{c_H}{2}\right) h Z^\mu Z_\mu.
\end{equation}

It is often helpful to parameterize the new physics effects in terms of modifications in the Higgs couplings, with \(\delta \kappa_3 \equiv \frac{g_{hhh}}{g_{hhh}^{\text{SM}}} - 1\) and \(\delta c_Z \equiv \frac{g_{hZZ}}{g_{hZZ}^{\text{SM}}} - 1\), which are related to \(c_6\) and \(c_H\) by:
\begin{equation}
    \delta \kappa_3 = \frac{c_6}{\lambda_\text{SM}} - \frac{3}{2} c_H, \quad \delta c_Z = -\frac{1}{2} c_H.
\end{equation}

The measurements of the Higgsstrahlung process (\(e^+e^- \to hZ\)) at future lepton colliders provide a powerful probe of \(\delta c_Z\). \(\delta \kappa_3\) contributes to this process only at the loop level. However, this contribution is relevant for probing \(\delta \kappa_3\) due to the high precision of the Higgsstrahlung process \cite{McCullough:2013rea, Degrassi:2016wml, DiVita:2017vrr, Chiu:2017yrx}. At leading order in both \(\delta c_Z\) and \(\delta \kappa_3\), the cross-section of this process is given by:
\begin{equation}
    \frac{\sigma_{hZ}}{\sigma_{hZ}^{\text{SM}}} \approx 1 + 2 \delta c_Z + \left[C_1(\sqrt{s}) + 2\delta Z_H\right] \delta \kappa_3,
\end{equation}
where the \(2 \delta c_Z\) term comes from the tree-level Higgs wavefunction renormalization, \(C_1(\sqrt{s})\) captures the interference effect between the tree-level and one-loop amplitudes, and \(2\delta Z_H\) denotes the universal contribution from the one-loop Higgs self-energy. The energy dependence of \(C_1\) for the associated production of Higgs at lepton colliders is provided in Appendix A of Ref.~\cite{DiVita:2017vrr}. A single \(hZ\) measurement at a particular energy scale is insufficient to determine \(\delta c_Z\) and \(\delta \kappa_3\) separately. However, when multiple center-of-mass energies are delivered at a future collider, the energy dependence in \(C_1\) lifts the degeneracy between \(\delta c_Z\) and \(\delta \kappa_3\), allowing them to be simultaneously constrained.

At higher energies, direct multiple-Higgs-boson production becomes possible, directly constraining \(\delta \kappa_3\) at the tree level. For instance, when the ILC operates at a center-of-mass energy of \(\sqrt{s} = 500\;\text{GeV}\) with an integrated luminosity of \(\mathcal{L} = 4\;\text{ab}^{-1}\) \cite{Barklow:2015tja}, the collider can access both the single-Higgs production channel and the di-Higgs channel to probe \(c_6\) and \(c_H\) \cite{DiVita:2017vrr, deBlas:2019rxi}. According to Ref.~\cite{deBlas:2019rxi}, the \(1\sigma\) constraint on \(|\delta \kappa_3| < 27\%\) due to the additional di-Higgs channel shrinks and tilts the exclusion contour for the ILC in \cref{fig:WilsonCoef}.

On the high-energy hadron collider side, the HL-LHC serves as the immediate future collider to provide constraints on trilinear couplings by measuring the di-Higgs channel. It is expected to constrain \(|\delta \kappa_3| \lesssim 50\%\) at 68\% confidence level at \(\sqrt{s} = 14\;\text{TeV}\) with \(\mathcal{L} = 3\;\text{ab}^{-1}\) of data \cite{Cepeda:2019klc}. The constraints on the trilinear coupling from di-Higgs channels at future hadron colliders are discussed in studies on FCC-hh \cite{Contino:2016spe, Banerjee:2018yxy}, with the estimated precision on the Higgs self-coupling at \(\sim \mathcal{O}(5\%)\) at \(\sqrt{s} = 100\;\text{TeV}\) with \(\mathcal{L} = 30\;\text{ab}^{-1}\). Hence, we consider the possibility that both FCC-ee and FCC-hh will be delivered in the future, further tightening their joint constraints on \(c_6\) and \(c_H\) as multi-Higgs production channels open up.

Furthermore, a future high-energy muon collider can also constrain \(c_6\) and \(c_H\). With \(\sqrt{s} = 10\;\text{TeV}\) and \(\mathcal{L} = 10\;\text{ab}^{-1}\) as the operational benchmark \cite{AlAli:2021let, Black:2022cth}, \(\delta c_Z\) can be constrained to \(\sim \mathcal{O}(0.1\%)\) by a global fitting over various Higgs precision measurements at a muon collider \cite{Li:2024joa}, and \(\delta \kappa_3\) can potentially be excluded at the level of \(\sim \mathcal{O}(5\%)\) \cite{Han:2020pif}. With these inputs, we provide an exclusion contour for a 10-TeV muon collider in \cref{fig:WilsonCoef}.

\begin{table}
    \centering
    \begin{tabular}{l|cr|cr|ccr}
        \hline \hline
         & \multicolumn{2}{c|}{CEPC} & \multicolumn{2}{c|}{FCC-ee} & \multicolumn{3}{c}{ILC} \\ 
         \hline
        $\sqrt{s}\;(\text{GeV})$  & $240$ & \multicolumn{1}{c|}{$360$} & $240$ & \multicolumn{1}{c|}{$365$} & $250$ & \multicolumn{1}{c}{$350$} & \multicolumn{1}{c}{$500$} \\
        $\mathcal{L}\;(\text{ab}^{-1})$ & \multicolumn{1}{r}{$20$}  & $1$ & \multicolumn{1}{r}{$5$} & $1.5$ & \multicolumn{1}{r}{$2$} & \multicolumn{1}{r}{$0.2$} & $4$ \\
        $\Delta \sigma / \sigma$ & \multicolumn{1}{r}{0.26\%} & 1.4\% & \multicolumn{1}{r}{0.5\%} & 0.9\% & \multicolumn{1}{r}{0.71\%} & 2.1\% & 1.1\% \\ 
        \hline \hline
    \end{tabular}
\caption{Projected precision for the cross section of associated production \(e^+e^- \to hZ\) for various future collider benchmarks \cite{CEPCPhysicsStudyGroup:2022uwl, Bernardi:2022hny, ILCInternationalDevelopmentTeam:2022izu}. Note that while plotting \cref{fig:WilsonCoef}, ILC has an additional constraint from the di-Higgs channel when \(\sqrt{s} = 500\;\text{GeV}\).}
\label{tab:FutureColliderBenchmarks}
\end{table}

\section{Unitarity Bounds on UV Parameters of the Singlet Extension \label{app:UnitarityBounds}}

This section provides details on the unitarity bounds obtained for the singlet model, whose most general UV Lagrangian is shown in \cref{eqn:SingletGeneral}. For our study, we considered the $2$-to-$2$ tree-level $s$-wave perturbative unitarity. In this case, the pair of both the initial-state particles and the final-state particles are back-to-back, respectively, and the dynamics is described by the center-of-mass energy $\sqrt{s}$ and the scattering angle $\theta$. One can then perform a partial wave decomposition for the $2$-to-$2$ scattering amplitudes $\mathcal{A}$ by
\begin{equation}
    \mathcal{A}(s,t) = 16\pi \sum_{j=0}^{\infty} (2j+1) P_j(\cos\theta) \sqrt{S_i S_f} \cdot \sqrt{\frac{s}{4 |\mathbf{p}_i| |\mathbf{p}_f|}} a_j(s),
\end{equation}
where $P_j(\cos\theta)$ is the Legendre polynomial, $S_{i(f)}$ denotes the initial-(final-)state phase space symmetry factor, $|\mathbf{p}_{i(f)}|$ is the magnitude of the three-momentum of the initial-(final-)state particle in the center-of-mass frame, and $a_j(s)$ is the $j$\textsuperscript{th} partial wave coefficient. For each partial wave, the unitarity circle dictates that $|\Re(a_j)| \leq 1/2$. Under the assumption that $\sqrt{s}$ is significantly larger than the masses of the external particles, one finds that the $s$-wave partial wave coefficients are related to the scattering amplitude $\mathcal{A}$ in a simple way
\begin{equation}
    a_0^{i \to f} = \lim_{s \to \infty} \frac{1}{16\pi s\sqrt{S_i S_f}} \int_{-s}^0 \mathrm{d}t \; \mathcal{A}^{i \to f}(s,t).
\end{equation}
To further simplify the computation, we note that for the singlet extension, only scalar exchanges will be of interest. As $\sqrt{s}$ grows, the most divergent part of the amplitude comes from the 4-point contact interaction, while those involving 3-point interactions will be suppressed by the propagator $\sim \mathcal{O}(1/s)$.

For our study, we considered the following scattering processes: $hh \to hh$, $r_s r_s \to r_s r_s$, $hh \to r_s r_s$, and $h r_s \to h r_s$, where $h$ denotes the Higgs boson and $r_s$ denotes the radial mode of the singlet after shifting to its VEV. First, note that the quartic interaction between the radial mode $r_s$ and the Higgs boson $h$ can only come from $|H|^4$, $|H|^2 S^2$, and $S^4$ terms in the UV Lagrangian, regardless of whether or not $H$ and $S$ have VEVs. Hence, the 4-point interaction between $h$ and $r_s$ is described by the Lagrangian:
\begin{equation}
    \mathcal{L} \supset -\frac{6\lambda_H}{4!} h^4 - \frac{\lambda_s}{4!} r_s^4 - \frac{\lambda_m}{4} h^2 r_s^2. 
\end{equation}
This translates to the following matrix of partial wave coefficients:
\begin{equation}
    \begin{pmatrix} 
    hh \to hh & hh \to r_s r_s \\
    r_s r_s \to hh & r_s r_s \to r_s r_s, h r_s \to h r_s
    \end{pmatrix} = \frac{1}{2} 
    \begin{pmatrix} 
    \frac{3 \lambda_H}{8\pi} & \frac{\lambda_m}{16\pi} \\
    \frac{\lambda_m}{16\pi} & \frac{\lambda_s}{16\pi}, \frac{\lambda_m}{8\pi}
    \end{pmatrix}.
\end{equation}
Hence, demanding that each scattering channel satisfies the unitarity bound translates to:
\begin{equation}
    \lambda_s \leq 16\pi, \quad \lambda_H \leq \frac{8\pi}{3}, \quad \lambda_m \leq 8\pi.
\end{equation}
Also, when treating the scattering matrix $S$ as a unitary operator, one would also demand that the maximal eigenvalue of the matrix of partial wave coefficients is less than $1/2$. This translates to:
\begin{equation}
    6 \lambda_H + \lambda_s + \sqrt{(6 \lambda_H - \lambda_s)^2 + 4 \lambda_m^2} \leq 32\pi.
\end{equation}
Lastly, the tree-level renormalization condition that sets the Higgs quartic coupling in the EFT to \(\lambda_\text{SM}\) implies that:
\begin{equation}
    \lambda_H = \frac{m_h^2}{2v_\text{SM}^2} + \frac{\left[ \lambda_m \left(3A_s + \sqrt{9A_s^2 + 24\lambda_s}\right) - 2A_m \lambda_s \right]^2}{2\lambda_s \left[ A_s \left(3A_s + \sqrt{9A_s^2 + 24\lambda_s}\right) + 8 \lambda_s \right]}.
\end{equation}
These are all the constraints pertinent to the perturbative unitarity bound presented in \cref{eqn:ParamConstraints}.

\bibliographystyle{utphys}
\bibliography{references}

\end{document}